\documentclass[submission,copyright,creativecommons]{eptcs}
\providecommand{\event}{ACL2 Workshop 2015} 
\usepackage{breakurl}             
\usepackage{acl2}

\providecommand{\urlalt}[2]{\href{#1}{#2}}
\providecommand{\doi}[1]{doi:\urlalt{http://dx.doi.org/#1}{#1}}

\newcommand{\ptt}[1]{\tt{#1}}

\title{Stateman:  Using Metafunctions to Manage Large Terms Representing Machine States}
\author{J Strother Moore
\institute{Department of Computer Science\\
The University of Texas at Austin}
\email{moore@cs.utexas.edu}
\thanks{This work was partially supported by ForrestHunt, Inc.}
}

\def\titlerunning{Stateman}
\def\authorrunning{J S.~Moore}
\begin{document}
\maketitle

\begin{abstract}

When ACL2 is used to model the operational semantics of computing machines,
machine states are typically represented by terms recording the contents of
the state components.  When models are realistic and are stepped through
thousands of machine cycles, these terms can grow quite large and the cost of
simplifying them on each step grows.  In this paper we describe an ACL2 book
that uses {\ptt{HIDE}} and metafunctions to facilitate the management of
large terms representing such states.  Because the metafunctions for each
state component updater are solely responsible for creating state expressions
(i.e., ``writing'') and the metafunctions for each state component accessor
are solely responsible for extracting values (i.e., ``reading'') from such
state expressions, they can maintain their own normal form, use {\ptt{HIDE}}
to prevent other parts of ACL2 from inspecting them, and use honsing to
uniquely represent state expressions.  The last feature makes it possible to
memoize the metafunctions, which can improve proof performance in some
machine models.  This paper describes a general-purpose ACL2 book modeling a
byte-addressed memory supporting ``mixed'' reads and writes.  By ``mixed'' we
mean that reads need not correspond (in address or number of bytes) with
writes.  Verified metafunctions simplify such ``read-over-write'' expressions
while hiding the potentially large state expression.  A key utility is a
function that determines an upper bound on the value of a symbolic arithmetic
expression, which plays a role in resolving writes to addresses
given by symbolic expressions.  We also report on a preliminary experiment
with the book, which involves the production of states containing several
million function calls.


\end{abstract}

\section{Background}

ACL2 \cite{car,acs} is frequently used to model computing machines via
operational semantics.  It is not difficult to configure the ACL2 theorem
prover so that it can use the definitions of the machine semantics and a few
well-chosen rewrite rules to step through code sequences, split on tests,
induct on loops, etc.  Examples of these methods being used to prove
functional correctness of code under formal operational semantics may be
found in numerous publications
\cite{Liu-04,Moore-Martinez09,Toibazarov,fmcad14}.  Such symbolic state terms
can grow quite large when many steps are composed.  The question addressed
here is: {\em{how can we exploit ACL2's rewriter to symbolically execute
    formalized code while preventing it from slowing down as state
    expressions get large?}}

This paper describes the Stateman book for managing large terms representing
machine states in ACL2 models of computing machines.  ``Stateman'' stands for
``state management.''  This is a work in progress and this paper has many
brief descriptions of intended {\bf{Future Work}}.

The idealistic dream is that a user wishing to model some byte-addressed
computing machine and do code proofs or run the Codewalker
tool\footnote{Codewalker extracts ACL2 functions from machine code given the
  formal operational semantics of the ISA and is similar to the HOL
  decompilation work by Magnus Myreen\cite{Myreen-09,Myreen-12}.  See the
  {\ptt{README}} file in the Community Book directory
  {\ptt{projects/codewalker/}}.  The version of Codewalker used here is still
      experimental.} might build the operational semantics on top of the
state provided by Stateman and thereby inherit the state management
techniques here described.  But machine models are very idiosyncratic.  Users
may actually need to design their own states and merely exploit the basic
techniques described here.  Thus, this paper focuses mainly on the design
decisions in our work.  As usual, readers are welcome, indeed encouraged, to
read the Stateman book itself and use it as the basis of their own versions.


We start with a brief description of our generic state, then we present the
highlights of our state management techniques, provide some examples, discuss
a few details, and present some preliminary performance measures.

\section{The Generic State}

The book provides a generic single-threaded object, {\ptt{ST}} (henceforth,
$st$), providing three fields.  See :DOC stobj.\footnote{When we say ``See
  :DOC $x$'' we mean see the documentation topic $x$ in the ACL2
  documentation, which may be found by visiting the ACL2 home
  page\cite{acl2-home-page}, clicking on {\underline{The User's Manuals}},
  then clicking on the {\underline{ACL2+Books Manual}} and typing $x$ into
  the ``Jump to'' box.}

\begin{acl2p}
(defstobj st
  (I  :type unsigned-byte :initially 0)          ; program counter
  (S  :initially nil)                            ; status
  (M  :type (array (unsigned-byte 8) (*m-size*)) ; memory
      :initially 0
      :resizable nil
      )
  :inline t
  :renaming
  ((UPDATE-I   !I)
   (UPDATE-S   !S)
   (UPDATE-MI  !MI)
   (M-LENGTH   ML)))
\end{acl2p}

The primitive accessors are {\ptt{I}}, {\ptt{S}}, and {\ptt{MI}}, and the
primitive updaters are {\ptt{!I}}, {\ptt{!S}}, and {\ptt{!MI}}.\footnote{The third field of
the single-threaded object is named {\ptt{M}} and is an array, but only the elements can be
accessed or changed, with {\ptt{MI}} and {\ptt{!MI}}.}   The
{\ptt{I}} and {\ptt{S}} fields were originally intended for the machine's
instruction counter and status flag, and {\ptt{MI}} provides a byte addressed
memory of 8-bit bytes.  The person using this book to model the state of a
computing machine need not use the {\ptt{I}} and {\ptt{S}} fields for their
implied purposes.  The modeler might, for instance, choose to store all state
information including the instruction counter and various status bits in the
byte addressed memory and ignore the {\ptt{I}} and {\ptt{S}} fields
altogether.

Byte-addresses are integers starting at 0.  The byte-addressed memory is of
fixed size, {\ptt{*m-size*}}, which is currently only {\ptt{5312}}.  This
constant is a holdover from the earliest use of the state and ({\bf{Future
    Work}}) will be generalized in future work.  Indeed, the whole
development would have been easier were there no upper bound on memory size.  Imposing
an upper bound forced certain issues to be dealt with -- issues that are necessarily raised in
any realistic model.  The magnitude of that upper bound is practically irrelevant from
the research perspective.

The Stateman book uses {\ptt{MI}} and {\ptt{!MI}} only to provide support for
two more general utilities, {\ptt{R}} and {\ptt{!R}}, for reading and writing
an arbitrary number of bytes.  We do not think of {\ptt{MI}} and {\ptt{!MI}}
as ``visible'' to the user of Stateman.

It is best to think of the generic state as providing the following
functionality:

\begin{tabular}{ll}
expression &  value \\
{\ptt{(I $st$)}} &       instruction counter of state $st$\\
{\ptt{(S $st$)}} &      status flag of state $st$\\
{\ptt{(R $a$ $n$ $st$)}} & natural number obtained by reading $n$ bytes starting\\
                           & at address $a$ in the memory of state $st$\\
{\ptt{(!I $v$ $st$)}} &    new state obtained from state $st$ by setting\\
                      &    the instruction counter to $v$\\
{\ptt{(!S $v$ $st$)}} &    new state obtained from state $st$ by setting the status\\
                      &    flag to $v$\\
{\ptt{(!R $a$ $n$ $v$ $st$)}} & new state obtained by writing $n$ bytes of natural\\
                      & number $v$ into the memory of $st$ starting at address $a$\\
\end{tabular}

{\ptt{R}} and {\ptt{!R}} use the ``Little Endian'' convention.  For example,
{\ptt{(!R $a$ $n$ $v$ $st$)}} writes the less significant bytes of $v$ to the lower
addresses, with the least significant byte written to address $a$ and all
other bytes written to larger addresses.  ({\bf{Future Work}}) We would like
to support either Little or Big Endian conventions.

Nests of {\ptt{!I}}, {\ptt{!S}}, and {\ptt{!R}} applications are called
{\em{state expressions}} or {\em{state terms}} because they denote machine
states.  Any term whose top function symbol is {\ptt{I}}, {\ptt{S}}, or
{\ptt{R}} applied to a state expression is called a {\em{read-over-write}}
expression.  Any term whose top function symbol is {\ptt{!I}}, {\ptt{!S}}, or
{\ptt{!R}} applied to a state expression is called a {\em{write-over-write}}
expression.  Of course, write-over-write expressions are themselves state
expressions.  

Our concern here is simplifying read-over-write and write-over-write
expressions in support of code proofs and code walks.  These issues are
straightforwardly managed with rewrite rules.  For example, the read over
write expression {\ptt{(R 24 8 (!R 40 8 $v$ $st$))}} can be simplified to
{\ptt{(R 24 8 $st$)}}. But as state expressions grow large -- and they
can grow very large when long code sequences are involved -- two problems
crop up.

First, the rewriter tends to re-simplify parts of states that have already
been simplified.  Second, the traditional rewrite rules for handling byte-addressed memory
involve backchaining to establish that byte sequences do not overlap.  For
example, the rewrite rules that replace {\ptt{(R $a$ $n$ (!R $b$ $k$ $v$
    $st$))}} by {\ptt{(R $a$ $n$ $st$)}} have the hypotheses {\ptt{(natp
    $a$)}}, {\ptt{(natp $b$)}}, {\ptt{(natp $n$)}}, {\ptt{(natp $k$)}}, and
either {\ptt{(< (+ $a$ $n$) $b$)}} or {\ptt{(< (+ $b$ $k$) $a$)}}.  The
inequalities can get very expensive when $a$ and $b$ are large arithmetic
expressions.  Furthermore, $a$ and $b$ typically become large arithmetic
expressions when the code being explored is doing indexed addressing (as in
array access) and long code sequences are involved in the computation of the
indices.  Every read-over-write and write-over-write expression raises such
an {\em{overlap}} question.  Furthermore, a read of a deeply nested state
expression typically raises an overlap question for each write in the nest.
For speed we must answer overlap questions without resorting to heavy-duty arithmetic.

\section{Highlights of Key Design Decisions}

Some of the key decisions in the design of Stateman are listed and briefly
elaborated below.  In the next section, where we give examples, we discuss
the implications of some of these decisions.

\begin{itemize}

\item {\bf{Manage read-over-write and write-over-write expressions
    exclusively with metafunctions}}: Stateman defines a metafunction for
  each of {\ptt{I}}, {\ptt{S}}, {\ptt{R}}, {\ptt{!I}}, {\ptt{!S}}, and
  {\ptt{!R}}.  These metafunctions are named {\ptt{meta-I}}, {\ptt{meta-S}},
  etc.  Like all metafunctions, they take terms as input and yield possibly
  different terms as output.\footnote{Metafunctions traffic in fully
    translated terms but the examples in this paper generally show
    untranslated terms for readability.}  The metafunctions for {\ptt{R}} and
  {\ptt{!R}} are extended metafunctions and thus additionally take the
  metafunction context and ACL2 state as arguments.  These two metafunctions
  only use the type-alist in the metafunction context and they ignore the
  ACL2 state.  However, the biggest problem faced by these functions is the
  read-over-write overlap questions: ``is one address less than another?'', given
  only the syntactic expressions representing the two addresses.  This
  motivates the next item.

\item {\bf{Implement a syntactic interval inference mechanism}}: Imagine a
  function that when given an arithmetic/logical term, can infer an upper
  bound.  This is quite different functionality than normally found in ACL2.
  ACL2 can be configured to answer questions like ``Is $\alpha$ less than
  16?'' but here we want a utility for answering ``What number is $\alpha$
  less than?''  This functionality is especially important in codewalking
  unknown code.  Suppose the code in question uses $\alpha$ as an index into
  some array at location $base$.  What part of the state is changed if the
  code writes to $base+\alpha$?  If you know enough about the code to know
  the bound on the array, you could undertake to prove that $\alpha$ is in
  bounds and thus conclude that only the array is affected by the write.  But
  if you do not know much about the code, you need an inference mechanism to
  deduce a bound on $\alpha$.  Stateman provides a verified interval
  inference mechanism named {\ptt{Ainni}} which is discussed in more detail
  in Section \ref{ainni}.

\item {\bf{Implement syntactic means of deciding some inequalities}}: Given
  {\ptt{Ainni}}, it is possible to implement the extended metafunction
  {\ptt{meta-<}} that takes an inequality and the metafunction context and
  decides many inequalities, {\ptt{(< $\alpha$ $\beta$)}}, by computing
  intervals for $\alpha$ and $\beta$ and comparing their endpoints, e.g., if
  the upper bound of $\alpha$ is below the lower bound of $\beta$, then the
  inequality is true.  This can save backchaining into linear arithmetic on
  large arithmetic/logical expressions.

\item {\bf{Implement syntactic means of simplifying some {\ptt{MOD}}
    expressions}}: In machine arithmetic, expressions of the form {\ptt{(MOD
    $\alpha$ '$n$)}} frequently arise, where {\ptt{n}} is some natural
  number.  Some expressions of this sort can be simplified by syntactic means given
  the ability to infer bounds on $\alpha$.  See Section \ref{metamod}.

\item {\bf{Use syntactic means to decide overlap questions}}: Suppose the type-alist
  tells us that the 32-bit word at address 8, i.e., {\ptt{(R 8 4 $st$)}} is
  less than 16.  Then a quick syntactic scan of the address expression
  {\ptt{(+ 3200 (* 8 (R 8 4 $st$)))}} reveals that the value lies in the
  interval [3200, 3320] and so reading, say, 3 bytes from that address might
  touch any address in the interval [3200, 3322].

\item {\bf{Insist that all byte counts be quoted constants}}: This facilitates the
  interval analysis mentioned above.  We do not regard it as a restriction
  given Stateman's intended application for code analysis.  In most ISAs the
  number of bytes to be manipulated by an instruction is explicitly given in
  the instruction or else is fixed by the instruction or the architecture.

\item {\bf{Do not put nested {\ptt{!R}}-expressions into address order}}: We
  leave the most recent writes at the top of the state expression under the
  assumption that program code tends to read from addresses recently written.

\item {\bf{Eliminate perfectly shadowed writes}}: When {\ptt{!R}}, with address $a$ and
  byte count $n$, is applied to a state expression already containing an application of
  {\ptt{!R}} with address $a$ with byte count $n$, Stateman eliminates the inner (earlier) one.
  Similar considerations apply to nested {\ptt{!I}} and {\ptt{!S}} calls.
  This reduces the size of the final state expression.  But Stateman does not
  try to eliminate partially shadowed writes.  We explain below.

\item {\bf{Use {\ptt{hons}} rather than {\ptt{cons}} to create state expressions}}:
  This means that if the same state expression is created along different paths of
  a code proof or walk, no additional space is allocated; furthermore,
  hons facilitates the use of memoization.

\item {\bf{{\ptt{HIDE}} the state expressions produced by the
    metafunctions}}: This ensures that no rewrite rule touches them.  For
  example, if a machine model mentions an expression like
\begin{acl2p}
(!R 32 4 $v$
    (!R 8 4 (+ (R 8 4 $st$) 4)
        (!I 123
            (!S NIL $st$))))
\end{acl2p}
  as would happen if it set the status flag to {\ptt{NIL}}, the instruction
  pointer to {\ptt{123}}, incremented the word at address 8 by 4 and wrote
  $v$ to the word at address 32, then the inside-out rewriting of ACL2 would
  invoke the metafunctions for {\ptt{!S}}, then {\ptt{!I}}, and then
  {\ptt{!R}} (twice) and ripple a {\ptt{HIDE}} out so the final term would be
  as exactly as above but with a single {\ptt{HIDE}} around it at the top
  level.  It would never be further simplified except by these metafunctions.

\item{\bf{{\ptt{HIDE}} some values extracted by reads from hidden states to
    avoid re-simplifying them}}: This is a controversial decision and is
  still quite unsettled.  ({\bf{Future Work}}) The issue is that over long
  codewalks (involving thousands of instructions) the expressions built up as
  values in the memory can be huge.  By embedding extracted values in
  {\ptt{HIDE}} expressions, they are not re-simplified.  The downside is that
  it can be impossible to decide simple tests because one does not know much
  about the hidden expressions.  A compromise would be to bury the
  {\ptt{HIDE}}s several levels down in the extracted expressions, leaving the
  top few function symbols available.  At the moment, all extracted values
  are hidden except constants and calls of {\ptt{R}}.  This means that the
  metafunctions here must remove some {\ptt{HIDE}}s from values before
  storing them into memory.

\item {\bf{Prove guards and well-formedness guarantees of the
    metafunctions}}: ACL2 users should be well aware of the efficiency
  advantages of verifying the guards on functions used in heavy-duty
  computations. A less familiar topic, though, is discussed in the new
  feature documented in :DOC well-formedness-guarantee.  It has long been the
  case that when a metafunction is applied the theorem prover checks that the
  result is a well-formed term, by running the function {\ptt{termp}} on the
  output and the current ACL2 world.  This hidden cost of metafunctions goes
  all the way back to the origin of ACL2 in 1989.  However, when the output
  of a metafunction is huge, the well-formedness check can be expensive, and
  the basic supposition in the Stateman work is that state expressions are
  huge.  A new feature of ACL2 Version 7.2 makes it possible to skip the
  well-formedness check by {\em{proving}} that the metafunction always
  returns a {\ptt{termp}}.  We have found that providing such well-formedness
  guarantees is worthwhile in Stateman.  See \cite{wf-guarantee}.
  We give some data on this below.

\end{itemize}

\section{Examples}

We illustrate these ideas with a few examples.  The reader may notice two odd
aspects to our examples.  The first is that most addresses illustrated are
quoted constants.  The second is that when non-constant expressions occur as
addresses the only variable involved is $st$ and it always occurs in a
primitive state accessor like {\ptt{(R $a$ $n$ $st$)}}.  We do not believe
these are serious constraints if Stateman is used for code analysis: Typical
code, especially binary machine code, refers to fixed addresses or offsets
from other addresses (as in array indexing and stack slots relative to some
stack or frame pointer in a register); ``variables'' are just the
contents of memory locations at such addresses.  However ({\bf{Future Work}})
it would not be difficult to support variable symbols provided the context
established natural number bounds on their values.

Examples (1)--(7) below are extracted verbatim from a session log that
started in a fresh ACL2 with the inclusion of the Stateman book.  Because
this is a work in progress, we keep the version number as part of the book
name right now.  This log started by including {\ptt{stateman22.lisp}} which
is included in the supplemental material.  The supplemental material also
includes {\ptt{simple-examples.lsp}}, a file (not a book) showing the actual
input forms for these and some other examples in this paper.  We hope those
forms can help the user who wishes to extend Stateman's functionality.


\begin{acl2p}
ACL2 !>(meta-!I '(!I '123 st))                                         ;{\rm{(1)}}
(HIDE (!I '123 ST))
\vspace*{0.0em}
ACL2 !>(meta-!R '(!R '0 '4 (R '16 '4 st) (HIDE (!I '123 ST)))          ;{\rm{(2)}}
                nil state)
(HIDE (!R '0 '4 (R '16 '4 ST) (!I '123 ST)))                           ;{\rm{($st'$)}}
\vspace*{0.0em}
ACL2 !>(meta-I '(I (HIDE (!R '0 '4 (R '16 '4 st)  (!I '123 ST)))))     ;{\rm{(3)}}
'123
\vspace*{0.0em}
ACL2 !>(meta-R '(R '0 '4 (HIDE (!R '0 '4 (R '16 '4 st) (!I '123 ST)))) ;{\rm{(4)}}
               nil state)
(R '16 '4 ST)
\vspace*{0.0em}
ACL2 !>(meta-R '(R '2 '2 (HIDE (!R '0 '4 (R '16 '4 st) (!I '123 ST)))) ;{\rm{(5)}}
               nil state)
(HIDE (ASH (R '16 '4 ST) '-16))
\vspace*{0.0em}
ACL2 !>(meta-R '(R '8 '4 (HIDE (!R '0 '4 (R '16 '4 st) (!I '123 ST)))) ;{\rm{(6)}}
               nil state)
(R '8 '4 ST)
\vspace*{0.0em}
ACL2 !>(meta-R '(R '2 '4 (HIDE (!R '0 '4 (R '16 '4 st) (!I '123 ST)))) ;{\rm{(7)}}
               nil state)
(HIDE (BINARY-+ (ASH (R '4 '2 ST) '16)
                (ASH (R '16 '4 ST) '-16)))
\end{acl2p}

In example {\rm{(1)}} we call the metafunction for {\ptt{!I}} on the term {\ptt{(!I '123 st)}},
just as the rewriter does when it encounters a {\ptt{!I}}-term.  The result is
a hidden state.  Notice that metafunctions traffic in fully translated terms.

In example {\rm{(2)}} we call the metafunction for {\ptt{!R}} on the {\ptt{!R}}-term
that writes the 4-byte value of {\ptt{(R '16 '4 ST)}} to location {\ptt{0}} in
the previously produced (now hidden) state.  Note that the metafunction for {\ptt{!R}}
takes two additional arguments, the metafunction context, in this case {\ptt{nil}}, and the ACL2 state object,
since {\ptt{meta-!R}} is an extended metafunction.  Again, nothing significant happens
except the new state is hidden.   Henceforth in this narrative we will refer to the
state produced by {\rm{(2)}} as $st'$.

In example {\rm{(3)}} we use the metafunction for {\ptt{I}} to extract the instruction counter
of $st'$.

In example {\rm{(4)}} we use the metafunction for {\ptt{R}} to read (4 bytes of) the
contents of address 0 in $st'$.  The result is exactly what was written in {\rm{(2)}}
because it was 4 bytes long.

In example {\rm{(5)}} we read the last two bytes of that previously written
quantity, that is, we read 2 bytes starting at address 2 in $st'$.  Two things
are noteworthy.  One is that it is reported as the 4-byte quantity that was
written in {\rm{(2)}}, shifted down by 16 bits.  The second is that it is hidden -- the
``controversial'' decision.

In example {\rm{(6)}} we read from an address above any affected by the write in $st'$.
The result is whatever was there in the original state $st$.

In example {\rm{(7)}} we read 4 bytes starting at address 2 in $st'$.  This is a
``mixed'' read in the sense that the result involves the last two bytes from
what was written at address 0 and the bytes that were at locations 4 and 5 of
the original state $st$.  It is expressed as a sum, with the latter bytes
shifted up.  Again, it is (controversially) hidden.

It is important to realize that all of these transformations are carried out
by verified metafunctions without involving rewrite rules, linear arithmetic,
or other heavy-duty theorem proving.  Consequently, these transformations
are very fast.

Since the {\ptt{I}} and {\ptt{S}} slots are unaffected by writes to memory
and do not involve addresses or overlap issues our examples below focus on
{\ptt{R}}- and {\ptt{!R}}-terms.

Henceforth, we will display untranslated terms for both input and output and will not
exhibit the calls of the relevant metafunction.  Instead, the reader should
understand that the notation ``$\alpha \Longrightarrow \beta$''
means that $\alpha$ is transformed to $\beta$ by the metafunction appropriate
for the top function symbol of $\alpha$.  Since both {\ptt{meta-R}} and
{\ptt{meta-!R}} take a metafunction context we make clear in the surrounding
narrative what the context is.  This only involves describing the governing
assumptions (as encoded in the type-alist).  Finally, instead of writing
something like ``$\alpha \Longrightarrow$ {\ptt{(IF $hyp$ $\beta$
    $\alpha$)}}'' we will generally write ``$\alpha
\Longrightarrow^\dagger \beta$'' and describe the side condition
$hyp$ generated by the metafunction in the accompanying narrative.  Recall
that before such an $\alpha$ is replaced by $\beta$ the rewriter must
establish $hyp$.

Given a metafunction context in which the type-alist is empty, we can
thus recap lines {\rm{(1)}}--{\rm{(7)}} above with:
\begin{acl2p}
(!I 123 $st$)                                                            ;{\rm{(1)}}
$\Longrightarrow$
(HIDE (!I 123 $st$))
\vspace*{0.0em}
(!R 0 4 (R 16 4 st) (HIDE (!I 123 $st$)))                                ;{\rm{(2)}}
$\Longrightarrow$
(HIDE (!R 0 4 (R 16 4 $st$) (!I 123 $st$)))
\vspace*{0.0em}
(I (HIDE (!R 0 4 (R 16 4 st)  (!I 123 $st$))))                           ;{\rm{(3)}}
$\Longrightarrow$
123
\vspace*{0.0em}
(R 0 4 (HIDE (!R 0 4 (R 16 4 st) (!I 123 $st$))))                        ;{\rm{(4)}}
$\Longrightarrow$
(R 16 4 $st$)
\vspace*{0.0em}
(R 2 2 (HIDE (!R 0 4 (R 16 4 st) (!I 123 $st$))))                        ;{\rm{(5)}}
$\Longrightarrow$
(HIDE (ASH (R 16 4 $st$) -16))
\vspace*{0.0em}
(R 8 4 (HIDE (!R 0 4 (R 16 4 st) (!I 123 $st$))))                        ;{\rm{(6)}}
$\Longrightarrow$
(R 8 4 $st$)
\vspace*{0.0em}
(R 2 4 (HIDE (!R 0 4 (R 16 4 st) (!I 123 $st$))))                        ;{\rm{(7)}}
$\Longrightarrow$
(HIDE (+ (ASH (R 4 2 $st$) 16)
         (ASH (R 16 4 $st$) -16)))
\end{acl2p}

Relatively little work is done on simplifying writes, aside from looking for
shadowed writes to be deleted.  For example, one might wonder at the simple
\begin{acl2p}
(!R 8 4 $v$ $st$)                                                          ;{\rm{(8)}}
$\Longrightarrow$
(HIDE (!R 8 4 $v$ $st$))
\end{acl2p}
{\noindent}since $v$ might be too big to fit in 4 bytes.  But instead of truncating $v$ on
write we do so on read:

\begin{acl2p}
(R 8 4 (HIDE (!R 8 4 $v$ $st$)))                                           ;{\rm{(9)}}
$\Longrightarrow$
(HIDE (MOD (IFIX $v$) 4294967296))
\end{acl2p}

Now let the metafunction context encode the assumption that {\ptt{(R 16 4
    $st$)}} is less than 16.  In the example below, we treat {\ptt{(R 16 4
    $st$)}} as an index into a QuadWord array (8-byte per entry) based at
address 3200.

\begin{acl2p}
(R (+ 3200 (* 8 (R 16 4 $st$))) 8                                       ;{\rm{(10)}}
   (HIDE (!R 3600 4 $v$ (!R 8 4 $w$ $st$))))
$\Longrightarrow^\dagger$
(R (+ 3200 (* 8 (R 16 4 $st$))) 8 $st$)
\end{acl2p}

\begin{acl2p}
(!R (+ 3200 (* 8 (R 16 4 $st$))) 8 $u$                                    ;{\rm{(11)}}
    (HIDE (!R 3600 4 $v$ 
              (!R 8 4 $w$
                  (!R (+ 3200 (* 8 (R 16 4 $st$))) 8 $x$
                      $st$)))))
$\Longrightarrow$
(HIDE (!R (+ 3200 (* 8 (R 16 4 $st$))) 8 $u$
          (!R 3600 4 $v$ 
              (!R 8 4 $w$ $st$))))
\end{acl2p}
The ``$\dagger$'' on the transformation in {\rm{(10)}} indicates that a side
condition was generated.  That side condition is {\ptt{(<= (R 16 4 $st$)
    15)}}, and it must be established before the replacement is made.
Establishing such side conditions should be trivial since they are extracted
from the type-alist in the metafunction context.  Given that condition, we
see that the 8-byte read at {\ptt{(+ 3200 (* 8 (R 16 4 $st$)))}} may only
touch bytes in the interval [3200, 3327].  We discuss this interval analysis
further below.  But because of it, the metafunction can determine that
neither of the two writes in the hidden state of {\rm{(10)}} is relevant since the 4
bytes starting at 3600 are above the target interval and 4 bytes starting at
8 are below it.

Interestingly, no side condition is necessary on transformation {\rm{(11)}}.  If
{\ptt{(R 16 4 $st$)}} is sufficiently large the new write at {\ptt{(+ 3200 (* 8
    (R 16 4 $st$)))}} {\em{might}} shadow out the write at 3600, but that
does not matter because the new write is added at the top of the expression
(chronologically after the write at 3600), so the answer above is correct.
And, regardless of the magnitude of {\ptt{(R 16 4 $st$)}}, the new write
shadows out the earlier one at the exact same address and the earlier write
can be dropped.

Our final example is contrived to show a mixed read that spans several
chronologically separated writes.  The empty metafunction context is
sufficient for this example.  We will ultimately read 8 bytes starting at
address 3.  But consider the writes that create the relevant memory.  The
write of 4 bytes of $v$ at address 2 is partially shadowed by the write of 4
bytes of $u$ at address 0.  The writes at 14 and 19 are irrelevant because we
only need bytes 3 through 10.  The first byte of our answer is the high
order byte of $u$ written at address 3.  The next two bytes are the two high
order bytes of $v$ at addresses 4 and 5.  Then we get 3 bytes from the original
$st$ at addresses 6, 7, and 8, and finally we get the two low order bytes
from $w$ at addresses 9 and 10.  We then assemble these 8 bytes using the
Little Endian notation and put the final sum into ACL2's term order.

\begin{acl2p}
(R 3 8                                                                ;{\rm{(12)}}
   (HIDE 
    (!R 14 5 $x$
        (!R 0 4 $u$ 
            (!R 19 8 $y$
                (!R 9 2 $w$
                    (!R 2 4 $v$ $st$)))))))
$\Longrightarrow$
(HIDE (+ (ASH (R 6 3 $st$) 24)
         (+ (MOD (ASH (IFIX $u$) -24) 256)
            (+ (ASH (MOD (IFIX $w$) 65536) 48)
               (ASH (MOD (ASH (IFIX $v$) -16) 65536) 8)))))
\end{acl2p}
({\bf{Future Work}}) We are dissatisfied with the normal form of expressions
denoting the results of mixed reads.  To be more precise, we do not have
enough experience with it yet to know whether it is sufficient for our
purposes.  The current implementation uses {\ptt{IFIX}} to convert terms to
integer form as required by basic rules for {\ptt{ASH}} (if syntactic
analysis cannot establish that the term returns an integer), uses {\ptt{MOD}}
to truncate unneeded higher order bits, and uses {\ptt{ASH}} to shift bits
into the right locations.  The question however is this: Suppose such an
expression is written to a memory location and then one must read a few bytes
from it.  The current metafunctions produce {\ptt{ASH/MOD}}-terms that could
be further simplified.  But given the controversial decision to {\ptt{HIDE}} the
complicated results of reads, that simplification should be done inside
{\ptt{meta-R}}.

Stateman does not produce normalized states for at least two reasons.
First, it does not put writes into address order.  Second it does not
eliminate partial shadows.  Why bother to eliminate partially shadowed
material if one can read out the answers if and when needed?  This
consideration is especially relevant since resolving a partial shadow
generally makes the state syntactically {\em{larger}}, e.g., to resolve the
shadowing of the write at 2 above one would replace {\ptt{(!R 2 4 $v$ $st$)}}
by the larger term {\ptt{(!R 4 2 (ASH (IFIX $v$ -16)) $st$)}}.  It is not
clear this is an improvement.  Furthermore, we suspect partial shadowing is
fairly rare compared to ``perfect shadowing'' where the $n$ bytes starting at
address $a$ are repeatedly reused for different $n$ byte values.

({\bf{Future Work}}) But the lack of normalization raises the question of
determining state equality.  Stateman does not support state equality at the
moment.  But the plan is to support it by a metafunction that announces the
equality of two states formed by different sequences of writes to the same
initial state by checking that every read of every byte written to either
state produces the same expression.

\section{Ainni:  Abstract Interpreter for Natural Number Intervals}
\label{ainni}

Perhaps the most important idea to come out of this work
so far is the development and verification of an ACL2 function that takes the
quotation of a term together with a type-alist and attempts to determine a
closed natural number interval containing the value of the term.  This
function is called {\ptt{Ainni}}, which stands for {\em{Abstract Interpreter
    for Natural Number Intervals}}.  {\ptt{Ainni}} can be thought of as a
``type-inference'' mechanism for a class of ACL2 arithmetic expressions,
except the ``types'' it deals with are intervals over the naturals.

{\ptt{Ainni}} explores terms composed of constants, the state $st$, and the
function symbols {\ptt{+}}, {\ptt{-}}, {\ptt{*}}, {\ptt{R}}, {\ptt{HIDE}},
{\ptt{MOD}}, {\ptt{ASH}}, {\ptt{LOGAND}}, {\ptt{LOGIOR}}, and
{\ptt{LOGXOR}}.\footnote{Several of these symbols are macros that expand into
  calls of function symbols that {\ptt{Ainni}} actually recognizes.}  ({\bf{Future Work}})
This set of function symbols was determined by seeing what functions were
introduced by the codewalk of a particularly large and challenging test
program: an implementation of DES.  Essentially, {\ptt{Ainni}} should support
all of the basic functions used in the semantics of the ALU operations of the
machine being formalized.  We therefore anticipate that the list here will
have to grow.

{\ptt{Ainni}} recursively descends through the term ``evaluating'' the
arguments of function calls -- only in this case that means computing
intervals for them -- and then applying bounders (see the discussion of ``bounders'' in :DOC
{\ptt{tau-system}}) corresponding to the function symbols to obtain an
interval containing all possible values of the function call.  At the bottom,
which in this case are calls of {\ptt{R}}, {\ptt{Ainni}} uses the type-alist
to try to find bounds on reads that are tighter than the syntactically
apparent $0 \leq$ {\ptt{(R $a$ $n$ $st$)}} $\leq 2^{8n}-1$.  ({\bf{Future
    Work}}) It is here, at the ``bottom'' of the recursion, that we could add
support for variable symbols or unknown function symbols.

For example, consider the quotation of the term
\begin{acl2p}
(+ 288 (* 8 (LOGAND 31 (ASH (R 4520 8 $st$) -3)))).
\end{acl2p}
In the absence of any contextual information, {\ptt{Ainni}} returns the
natural number interval [288,536].  The reasoning is straightforward: we know
that {\ptt{(R 4520 8 $st$)}} is a natural in the interval [0, $2^{64}-1$].
The tau-bounder for {\ptt{ASH}} tells us that shifting it right 3 reduces
that to [0, $2^{61}-1$], and then the tau-bounder for {\ptt{LOGAND}} tells us
that bitwise conjoining it with {\ptt{31}} shrinks the interval to [0,31].
Multiplying by {\ptt{8}} makes the interval [0, 248], and adding {\ptt{288}}
makes it [288, 536].


By default {\ptt{(R 4520 8 $st$)}} is known to lie in
[0,$2^{64}-1$], but the type-alist might restrict it to a smaller interval.
For example, it might assert that {\ptt{(R 4520 8 $st$)}} $<$ {\ptt{24}}, in which case
{\ptt{Ainni}} determines that the term above lies in the interval [288,304].

In addition to returning the interval, {\ptt{Ainni}} also returns a flag
indicating whether the term was one that {\ptt{Ainni}} could confine to a
bounded natural interval and a list of hypotheses that must be true for its
interval to be correct.  These hypotheses have two sources: (i) assumptions
extracted from the context and (ii) {\ptt{Ainni}}'s inherent assumptions
(such as a built-in assumption that no computed value is negative\footnote{We
  anticipate that any ISA employing Stateman's byte-addressed memory would
  use twos-complement arithmetic.}, which might translate to the hypothesis
{\ptt{(not (< $x$ $y$))}} if the term is {\ptt{(- $x$ $y$)}}).

Finally, {\ptt{Ainni}} is verified to be correct.  That is, the certification of
Stateman involves a proof of the formal version of:

\begin{quotation} Let $x$ be the quotation of an ACL2 term and $ta$ be a type-alist.
Let $flg$, {\ptt{($h_1$ $\ldots$ $h_k$)}} and [$lo$, $hi$] be the flag, hypotheses, and the
interval returned by {\ptt{Ainni}} on $x$ and $ta$.
Then if $flg$ is true:

\begin{itemize}
\item {\ptt{($h_1$ $\ldots$ $h_k$)}} is a list of quotations of terms,

\item $lo$ and $hi$ are natural numbers such that $lo \leq hi$, and

\item if {\ptt{($\mathcal{E}$ $h_i$ $a$)}} $=$
  {\ptt{T}} for each $1 \leq i \leq k$, then $lo \leq$ {\ptt{($\mathcal{E}$ $x$ $a$)}} $\leq hi$,
where $\mathcal{E}$ is an evaluator that recognizes the function symbols handled by {\ptt{Ainni}}.
\end{itemize}
\end{quotation}

{\ptt{Ainni}} is used in {\ptt{meta-R}} to handle the overlap questions that
arise.  In addition, it is used in {\ptt{meta-<}} to decide some inequalities and in {\ptt{meta-MOD}}
to simplify some {\ptt{MOD}} expressions.

Furthermore, {\ptt{Ainni}} is fast.  For example, in the codewalk of the DES algorithm,
one particular index expression is a nest of 382 function calls containing
every one of the function symbols known to {\ptt{Ainni}}.  Just for fun, here is
the expression, printed ``almost flat'' (without much prettyprinting):
\vspace{0.5em}
\begin{acl2p}{\tiny{
(LOGIOR
  (LOGAND 32 (ASH (MOD (ASH (LOGXOR (LOGIOR (ASH (ASH (LOGIOR (ASH (MOD (ASH (R 4520 8 ST) 0) 2) 31) (ASH (LOGAND 4026531840
    (R 4520 8 ST)) -1) (ASH (MOD (ASH (R 4520 8 ST) -27) 2) 26) (ASH (MOD (ASH (R 4520 8 ST) -28) 2) 25) (ASH (LOGAND 251658240 (R 4520
    8 ST)) -3) (ASH (MOD (ASH (R 4520 8 ST) -23) 2) 20) (ASH (MOD (ASH (R 4520 8 ST) -24) 2) 19) (ASH (LOGAND 15728640 (R 4520 8 ST))
    -5) (ASH (MOD (ASH (R 4520 8 ST) -19) 2) 14) (ASH (MOD (ASH (R 4520 8 ST) -20) 2) 13) (ASH (LOGAND 983040 (R 4520 8 ST)) -7) (ASH
    (MOD (ASH (R 4520 8 ST) -15) 2) 8)) -8) 24) (ASH (LOGIOR (ASH (MOD (ASH (R 4520 8 ST) -16) 2) 31) (ASH (LOGAND 61440 (R 4520 8 ST))
    15) (ASH (MOD (ASH (R 4520 8 ST) -11) 2) 26) (ASH (MOD (ASH (R 4520 8 ST) -12) 2) 25) (ASH (LOGAND 3840 (R 4520 8 ST)) 13) (ASH (MOD
    (ASH (R 4520 8 ST) -7) 2) 20) (ASH (MOD (ASH (R 4520 8 ST) -8) 2) 19) (ASH (LOGAND 240 (R 4520 8 ST)) 11) (ASH (MOD (ASH (R 4520 8
    ST) -3) 2) 14) (ASH (MOD (ASH (R 4520 8 ST) -4) 2) 13) (ASH (MOD (R 4520 8 ST) 16) 9) (ASH (ASH (R 4520 8 ST) -31) 8)) -8)) (R (+
    4376 (* 8 (R 4536 8 ST)) (* 8 (- (R 4528 8 ST)))) 8 ST)) -40) 256) -2))
  (ASH (MOD (ASH (MOD (ASH (LOGXOR (LOGIOR (ASH (ASH (LOGIOR (ASH (MOD (ASH (R 4520 8 ST) 0) 2) 31) (ASH (LOGAND 4026531840 (R 4520
    8 ST)) -1) (ASH (MOD (ASH (R 4520 8 ST) -27) 2) 26) (ASH (MOD (ASH (R 4520 8 ST) -28) 2) 25) (ASH (LOGAND 251658240 (R 4520 8 ST))
    -3) (ASH (MOD (ASH (R 4520 8 ST) -23) 2) 20) (ASH (MOD (ASH (R 4520 8 ST) -24) 2) 19) (ASH (LOGAND 15728640 (R 4520 8 ST)) -5) (ASH
    (MOD (ASH (R 4520 8 ST) -19) 2) 14) (ASH (MOD (ASH (R 4520 8 ST) -20) 2) 13) (ASH (LOGAND 983040 (R 4520 8 ST)) -7) (ASH (MOD (ASH
    (R 4520 8 ST) -15) 2) 8)) -8) 24) (ASH (LOGIOR (ASH (MOD (ASH (R 4520 8 ST) -16) 2) 31) (ASH (LOGAND 61440 (R 4520 8 ST)) 15) (ASH
    (MOD (ASH (R 4520 8 ST) -11) 2) 26) (ASH (MOD (ASH (R 4520 8 ST) -12) 2) 25) (ASH (LOGAND 3840 (R 4520 8 ST)) 13) (ASH (MOD (ASH (R
    4520 8 ST) -7) 2) 20) (ASH (MOD (ASH (R 4520 8 ST) -8) 2) 19) (ASH (LOGAND 240 (R 4520 8 ST)) 11) (ASH (MOD (ASH (R 4520 8 ST) -3)
    2) 14) (ASH (MOD (ASH (R 4520 8 ST) -4) 2) 13) (ASH (MOD (R 4520 8 ST) 16) 9) (ASH (ASH (R 4520 8 ST) -31) 8)) -8)) (R (+ 4376 (* 8
    (R 4536 8 ST)) (* 8 (- (R 4528 8 ST)))) 8 ST)) -40) 256) -2) 32) -1)
  (ASH (MOD (ASH (MOD (ASH (LOGXOR (LOGIOR (ASH (ASH (LOGIOR (ASH (MOD (ASH (R 4520 8 ST) 0) 2) 31) (ASH (LOGAND 4026531840 (R 4520
    8 ST)) -1) (ASH (MOD (ASH (R 4520 8 ST) -27) 2) 26) (ASH (MOD (ASH (R 4520 8 ST) -28) 2) 25) (ASH (LOGAND 251658240 (R 4520 8 ST))
    -3) (ASH (MOD (ASH (R 4520 8 ST) -23) 2) 20) (ASH (MOD (ASH (R 4520 8 ST) -24) 2) 19) (ASH (LOGAND 15728640 (R 4520 8 ST)) -5) (ASH
    (MOD (ASH (R 4520 8 ST) -19) 2) 14) (ASH (MOD (ASH (R 4520 8 ST) -20) 2) 13) (ASH (LOGAND 983040 (R 4520 8 ST)) -7) (ASH (MOD (ASH
    (R 4520 8 ST) -15) 2) 8)) -8) 24) (ASH (LOGIOR (ASH (MOD (ASH (R 4520 8 ST) -16) 2) 31) (ASH (LOGAND 61440 (R 4520 8 ST)) 15) (ASH
    (MOD (ASH (R 4520 8 ST) -11) 2) 26) (ASH (MOD (ASH (R 4520 8 ST) -12) 2) 25) (ASH (LOGAND 3840 (R 4520 8 ST)) 13) (ASH (MOD (ASH (R
    4520 8 ST) -7) 2) 20) (ASH (MOD (ASH (R 4520 8 ST) -8) 2) 19) (ASH (LOGAND 240 (R 4520 8 ST)) 11) (ASH (MOD (ASH (R 4520 8 ST) -3)
    2) 14) (ASH (MOD (ASH (R 4520 8 ST) -4) 2) 13) (ASH (MOD (R 4520 8 ST) 16) 9) (ASH (ASH (R 4520 8 ST) -31) 8)) -8)) (R (+ 4376 (* 8
    (R 4536 8 ST)) (* 8 (- (R 4528 8 ST)))) 8 ST)) -40) 256) -2) 2) 4))
}}\end{acl2p}

While the first argument of the {\ptt{LOGIOR}} is easy to bound the second and third are problematic. 
{\ptt{Ainni}} bounds the {\ptt{LOGIOR}} to [0,63] in less than one hundredth
of a second on a MacBook Pro laptop with a 2.6 GHz Intel Core i7 processor.

By the way, the second argument of the {\ptt{LOGIOR}} above actually lies in
[0,15] and the third in [0,16].  But proving those two bounds with, say,
{\ptt{arithmetic-5/top}}, takes about 33 seconds each, without {\ptt{Ainni}}
and {\ptt{meta-<}}.  But the main point is that {\ptt{Ainni}}
{\em{infers}} a correct bound.

\section{Syntactic Simplification of MOD Expressions}
\label{metamod}

Machine arithmetic introduces many {\ptt{MOD}} expressions in which the
second argument is constant.  Stateman provides the extended metafunction
{\ptt{meta-MOD}} that implements the following rules, where $i$, $j$, and $k$ are
natural constants.  The function also uses a concept called ``syntactic
integer'' realized by a function which takes the quotation of a term and
determines whether it is obviously integer valued.  For example, a sum
expression is a syntactic integer provided the two arguments are syntactic
integers, an {\ptt{ASH}} expression is a syntactic integer provided the first
argument is, and a {\ptt{LOGAND}} expression is a syntactic integer
regardless of the shape of the arguments.  In the rules below, $x$, $x_1, \ldots, x_j$
must be syntactic integer expressions.

\begin{itemize}
\item {\ptt{(MOD $x$ 0)}} = $x$

\item {\ptt{(MOD $i$ $k$)}} can be computed if both arguments are constants

\item {\ptt{(MOD (MOD $z$ $j$) $k$)}} = {\ptt{(MOD $z$ j)}}, if $j \leq k$
\item {\ptt{(MOD (MOD $x$ $j$) $k$)}} = {\ptt{(MOD $x$ $k$)}}, if $k$ divides $j$
\item {\ptt{(MOD (R $a$ $i$ $st$) $k$)}} = {\ptt{(R $a$ $i$ $st$)}}, if $256^i \leq k$
\item {\ptt{(MOD (+ $x_1$ $\ldots$ (MOD $x$ $j$) $\ldots$ $x_j$) $k$)}} = {\ptt{(MOD (+ $x_1$ $\ldots$ $x$ $\ldots$ $x_j$) $k$)}}, if $k$ divides $j$
\item {\ptt{(MOD $x$ $k$)}} = $x$, if {\ptt{Ainni}} claims the upper bound of $x$ is below $k$
\end{itemize}

The last rule is not only applied to the argument of {\ptt{meta-MOD}} but
also to the output of the second-to-last rule.

Some of these rules are built into {\ptt{arithmetic-5/top}} but in the
interests of speed, Stateman does not export {\ptt{arithmetic-5/top}} and
does much arithmetic simplification in its metafunctions.

\section{Some Details of Meta-R and Meta-!R}

The most complicated of the metafunctions are {\ptt{meta-R}} and
{\ptt{meta-!R}}, which use all of the functionality described above.  The
former is actually more complicated than the latter because the former deals
with mixed read-over-write.  We briefly discuss some design issues for these
two functions, starting with the simpler, {\ptt{meta-!R}}, but we urge the
interested reader to inspect the code in the Stateman book.

Since a successful application of {\ptt{meta-!R}} will transform {\ptt{(!R
    $a$ '$n$ $v$ (HIDE $st'$))}} into {\ptt{(HIDE (!R $a$ '$n$ $v$ $st'$))}}, we
must be careful not to fire the metafunction too soon: none of the subterms
will be rewritten again!  Thus {\ptt{meta-!R}} checks whether $a$ or $v$ contain
terms that might still be rewritten, e.g., embedded {\ptt{IF}}s, unexpanded
{\ptt{LAMBDA}} applications, or read-over-writes that have not yet been
resolved.  If such subterms are found, the metafunction does not fire and
{\ptt{(!R $a$ '$n$ $v$ (HIDE $st'$))}} continues to be subject to rewriting.

If we decide to fire, we remove all {\ptt{HIDE}}S in $a$ and $v$; remember they are
probably arithmetic/logical expressions formed by the semantics of an
instruction operating on data extracted from memory and thus
(controversially) hidden.  When we remove {\ptt{HIDE}}s we actually compute
the depth of the deepest {\ptt{HIDE}} first and then copy only that far into
the term so as to avoid re-copying a honsed term.

Then we dive through $st'$ looking for a perfect shadow of a write to $a$ of
$n$ bytes.  This is actually a little more complicated than just looking for
a deeper {\ptt{(!R $a$ $n$ $\ldots$)}} because the addresses may not be fully
normalized.  By using {\ptt{Ainni}} we can identify some non-identical
addresses that are semantically equivalent in the current context.  As we
dive through $st'$ looking for a shadowed assignment we also compute its
depth, so we can come back and delete it without further interval analysis.

Moving on to {\ptt{meta-R}}, the main complication is mixed read-over-write.
The question is, given {\ptt{(R $a$ '$n$ (!R $b$ '$k$ $v$ $st$))}}, does part of the
answer lie within $v$ or not?  {\ptt{Ainni}} can be used to handle many
general overlap questions but we prefer not to use {\ptt{Ainni}} if simpler
techniques apply.  For example, if both $a$ and $b$ are constants we can just
skip over this {\ptt{!R}} or extract the appropriate bytes from $v$ (remember
$n$ and $k$ are constants).  But more generally, we ask whether $a$ and $b$
are offsets from some common address, e.g., $a$ might be {\ptt{(+ 8 $sp$)}}
and $b$ might be {\ptt{(+ 16 $sp$)}} where $sp$ is some expression denoting,
say, the stack pointer.  While neither address is constant we can still
determine whether reading $n$ bytes from $a$ takes us into the region
written, by doing arithmetic on the two constant offsets ({\ptt{8}} and
{\ptt{16}} in this example) and the constants $n$ and $k$.  When no common
reference address can be found, we use {\ptt{Ainni}}.  Space does not permit
further description of mixed read-over-write and we urge the reader to see
the Stateman code.

Furthermore, space does not permit discussion of the proof issues.  But
correctness, guards, and well-formedness guarantees are all proved.
Probably the most interesting and difficult proofs concerned mixed
read-over-write and the validity of removing a deeply buried perfectly
shadowed write {\em{without}} being able to determine whether intervening
writes also shadow it, i.e., how do you justify transforming
\begin{acl2p}
(!R $a$ $n$ $v_1$
    (!R $b$ $k$ $w$
        (!R $a$ $n$ $v_2$ $st$)))
\end{acl2p}
to
\begin{acl2p}
(!R $a$ $n$ $v_1$
    (!R $b$ $k$ $w$ $st$))
\end{acl2p}
without knowing the relations between $a$, $n$, $b$ and $k$?  The
formalization of the general result we need is an inductively proved
{\ptt{LOCAL}} lemma, named {\ptt{LEMMA3}} in {\ptt{stateman22.lisp}},
establishing the correctness of a function that deletes a perfectly shadowed
write at an arbitrary depth.  {\ptt{LEMMA3}} is used in the proof of
{\ptt{META-!R-CORRECT}}.

\section{Memoization}

We have experimented with memoization of the metafunctions introduced by Stateman.
Memoization is theoretically useful in code proofs because the same symbolic
state might be produced on different paths through the code.  In addition,
the contents of the same addresses might be read multiple times from the same state.
On the other hand, memoization imposes an overhead and is thus not always worthwhile.

Memoization hits most often if all of the arguments are honsed rather than
consed.  For example, if {\ptt{f}} is memoized and one has typed {\ptt{(f '(a . b))}} at the top-level,
then the value of {\ptt{f}} on that cons pair is stored in the hash table for {\ptt{f}}.  But if one then
types {\ptt{(f (cons 'a 'b))}} the memoized answer is not found and {\ptt{f}} is recomputed.
In Common Lisp terms, the argument must be {\ptt{EQ}} not
{\ptt{EQUAL}}.  All of the state expressions produced by our metafunctions
are honsed and thus uniquely represented.  But this alone will not make
{\ptt{(memoize 'meta-R)}}, for example, particularly useful.

First, memoization cannot be applied to an extended metafunction because one
of the arguments is the ACL2 (live) state.  So {\ptt{meta-R}}, which takes
{\ptt{state}} as an argument (because it is a requirement of extended
metafunctions) but which ignores {\ptt{state}}, is defined in terms of a wrapper,
{\ptt{memoizable-meta-R}} which does not take {\ptt{state}} and which takes
only the type-alist from the metafunction context, not the whole context.

Second, the term argument of {\ptt{meta-R}} is of the form {\ptt{(R $a$ '$n$
    (HIDE $st'$))}} and typically came from simplifying some {\ptt{R}}-term
in the model.  The {\ptt{(HIDE $st'$)}} is honsed because it was produced by
one of our metafunctions.  But the rest of the term is not.  So we
{\ptt{hons-copy}} it before calling the wrapper.  These {\ptt{hons-copy}}s
are not as expensive as they may seem because the (very large) states and
values extracted from them are already honsed.

Third, we must similarly {\ptt{hons-copy}} the type-alist.

Thus, 
\begin{acl2p}
(defun meta-R (x mfc state)
  (declare (xargs :stobjs (state)
                  :guard (pseudo-termp x))
           (ignore state))
  (memoizable-meta-R (hons-copy x)
                     (hons-copy (mfc-type-alist mfc))))
\end{acl2p}

Experiments have indicated that it is not worthwhile memoizing
{\ptt{meta-I}}, {\ptt{meta-S}}, {\ptt{meta-!I}} or {\ptt{meta-!S}}: they are
too simple.  We have settled on:

\begin{acl2p}
(memoize 'memoizable-meta-r)
(memoize 'memoizable-meta-!r)
(memoize 'memoizable-meta-mod)
(memoize 'memoizable-meta-<)
\end{acl2p}

While {\ptt{Ainni}} is an obvious candidate for memoization, the functions
above include all of {\ptt{Ainni}}'s callers so it is not worthwhile.

Finally, when a metafunction fires -- even a metafunction with a
well-formedness guarantee -- the output is put into {\em{quote normal form}}
by which we mean all ACL2 primitives applied to constants are evaluated to
constants.  That is, {\ptt{(CONS '1 '2)}} is not in quote normal form, but
{\ptt{'(1 . 2)}} is.  This reduction to quote normal form is done by applying
the empty substitution to the term with the ACL2 utility {\ptt{sublis-var1}}.
We have found it worthwhile to memoize this function, but only when the
substitution is empty and the form being normalized is hidden (and thus
probably one produced by our metafunctions and thus honsed).

\begin{acl2p}
(memoize 'sublis-var1
         :condition '(and (null alist)
                          (consp form)
                          (eq (car form) 'HIDE)))
\end{acl2p}

({\bf{Future Work}})  More experimentation must be done before we are
comfortable with these decisions.  In addition, it might be practical to
make well-formedness guarantees ensure quote normal form.

\section{Preliminary Performance Results}

We have tested Stateman on only one very stressful example.  Roughly put the
setup for this example (which is not provided here) is as follows: Using the
state provided by Stateman, we defined an ISA for a register machine that
provides conventional but realistic arithmetic/logical functionality,
addressing modes, and control flow.  We then implemented a compiler from a
subset of ACL2 into this ISA.  After allocating declared arrays, constants,
etc., the compiler uses the rest of the memory to provide a call stack whose
stack and frame pointers are among the earlier addresses.  The compiler then
compiles a system of ACL2 functions and a main program as though it were
running on a stack machine, e.g., {\ptt{(LOGAND $x$ $y$)}} is compiled by compiling
$x$ and $y$ so as to leave two items on the stack, and then laying down a
block of code to pop those two items into temporary registers, apply the
{\ptt{LOGAND}} instruction to those registers, and push the result.
Addressing modes are used whenever possible to minimize the number of
instructions needed.  We then compiled an ACL2 implementation of the DES
algorithm.\footnote{Warren Hunt provided the definitions of the ISA and the
  DES algorithm in ACL2.}  The result is a code block of 15,361 instructions.
We then ran an experimental version of Codewalker on this code.

Using Codewalker and the state management techniques described here, ACL2
explores the code above and generates both clock and semantic functions for
DES.\footnote{As of this writing the Codewalker exploration of DES does not
  perform its standard ``projection'' (the transformation of functions that
  describe the entire state to functions that describe the contents of
  specific state components) because ACL2 gets a stack overflow trying to
  handle states of such large size.  ({\bf{Future Work}}) Clearly, additional
  work is necessary on Codewalker and/or ACL2 itself to handle the terms
  being produced by Stateman.}

The largest symbolic state in the decompilation of the DES algorithm
represents one path through the 5,280 instructions in the decryption loop.
The state contains 2,158,895 function calls consisting of one call of
{\ptt{!I}} and {\ptt{!S}} each and 58 calls of {\ptt{!R}} to distinct
locations.  That state expression also contains 459,848 calls of {\ptt{R}}
and 1,698,987 calls of arithmetic/logical functions such as {\ptt{+}}, and
{\ptt{*}}, {\ptt{LOGAND}}, {\ptt{LOGIOR}}, {\ptt{LOGXOR}}, {\ptt{ASH}}, and
{\ptt{MOD}}.  The values written are often very large.  The largest value
expression written is given by a term involving 147,233 function
applications, 31,361 of which are calls of {\ptt{R}} and the rest are calls
of arithmetic/logical functions.

We would like to be able to compare the performance of the current version of
Stateman to older techniques (in which rewrite rules alone are used to
canonicalize symbolic states) but Codewalker is unable to complete the
exploration of our implementation of DES using those older techniques.  The
time it takes to symbolically execute successive instructions increases
alarmingly, sometimes apparently exponentially (depending on the instruction
being executed) as the state sizes increase.  Of course, one might address
that with better rewrite rules, metafunctions, etc., but that was the origin
of the Stateman project.

However, we can provide some timing statistics on different versions of
Stateman.  The times shown are times taken to generate the clock and
semantics functions of our DES implementation on a MacBook Pro laptop with a
2.6 GHz Intel Core i7 processor with 16 GB of 1600 MHZ DDR3 memory.
Times are as measured by {\ptt{time\$}} and reported as ``realtime'' on a
otherwise unloaded machine.

Roughly put, guard verification saved 33 seconds, well-formedness guarantees
saved 337 more seconds, honsing as opposed to consing the metafunction
answers saved 124 more seconds even though no memoization was employed, and
memoizing then saved 119 more seconds.  Of particular interest is that
well-formedness guarantees were an order of magnitude more effective than
guard verification and that honsing even without memoization was a win
(presumably because less time was spent in allocation).

\vspace*{1em}

\begin{tabular}{ll}
without guard verification, well-formedness guarantees, & 988 seconds \\
honsing or memoization & \\
\\
with guard verification but without well-formedness & 955 seconds \\
guarantees, honsing, or memoization & \\
\\
with guard verification and well-formedness guarantees, & 618 seconds \\
but without honsing or memoization & \\
\\
with guard verification, well-formedness guarantees, & 494 seconds \\
and honsing, but without memoization & \\
\\
with guard verification, well-formedness guarantees, & 375 seconds \\
honsing, and the memoization described & \\
\end{tabular}

\vspace*{1em}

\section{Acknowledgments}

I especially thank Warren Hunt for his invaluable help during
the development of this software.  Warren developed the definitions and
proved many of the basic rewrite rules for {\ptt{I}}, {\ptt{S}}, {\ptt{R}},
{\ptt{!I}}, {\ptt{!S}}, and {\ptt{!R}}, as well as an ACL2 implementation of
DES and the formal semantics for the ISA to which the stack machine compiles.
I thank Matt Kaufmann, who gave me some strategic advice
on lemma development to prove the correctness of one of the metafunctions as
well as his usual extraordinary efforts to maintain ACL2 while I pursue
topics such as this one.  Finally, the reviewers of this paper improved it
significantly and I am grateful for their careful and constructive criticism.


\end{document}